\documentclass[12pt]{article} 
\usepackage{a4}
\usepackage{amsmath}
\usepackage{amssymb}
\usepackage{latexsym}
\usepackage{amssymb}
\usepackage{epsfig}
\usepackage{natbib}
\def \ii{{\mathrm{i}}}

\def \TP{{\mathrm{P}}}
\def \d{{\mathrm{d}}}

\def \pd{\partial}

\def \e{{\mathrm{e}}}

\def \Bx{{\boldsymbol{x}}}

\def \FF{{\cal{F}}}

\def \rr{{\boldsymbol{r}}}
\def \xx{{\boldsymbol{x}}}

\def \kk{{\boldsymbol{k}}}

\begin{document}
\title{{\bf The fundamentals of 
non-singular dislocations in the theory of 
gradient elasticity: dislocation loops and straight dislocations   
}}
\author{
Markus Lazar~$^\text{a,b,}$\footnote{
{\it E-mail address:} lazar@fkp.tu-darmstadt.de (M.~Lazar).} 
\\ \\
${}^\text{a}$ 
        Heisenberg Research Group,\\
        Department of Physics,\\
        Darmstadt University of Technology,\\
        Hochschulstr. 6,\\      
        D-64289 Darmstadt, Germany\\
${}^\text{b}$ 
Department of Physics,\\
Michigan Technological University,\\
Houghton, MI 49931, USA
}

\date{\today}    
\maketitle

\begin{abstract}
The fundamental problem of non-singular dislocations
in the framework of the theory of gradient elasticity
is presented in this work.
Gradient elasticity of Helmholtz type and bi-Helmholtz type are used. 
A general theory of non-singular dislocations is developed for 
linearly elastic, infinitely extended, homogeneous, and isotropic media.
Dislocation loops and straight dislocations are investigated. 
Using the theory of gradient elasticity, 
the non-singular fields which are produced 
by arbitrary dislocation loops are given.
`Modified' Mura, Peach-Koehler, and Burgers formulae 
are presented in the framework of gradient elasticity theory. 
These formulae are given in terms of an elementary function, which 
regularizes the classical expressions, obtained from the Green
tensor of the Helmholtz-Navier equation and bi-Helmholtz-Navier equation.
Using the mathematical method of Green's functions and the Fourier
transform, exact, analytical, and non-singular 
solutions were found.
The obtained dislocation fields are non-singular due to the 
regularization of the classical singular fields.
\\

\end{abstract}

\section{Introduction}
Dislocations play an important role in the understanding of many phenomena in
solid state physics, materials science, and engineering. 
They are the primary carriers of crystal plasticity.
Dislocations are line defects which can be straight or curved lines.
The internal geometry of generally curved dislocations, in deformed crystals is
very complex. 
In the classical theory of dislocation loops in isotropic materials~\citep{deWit60,Lardner,HL}  
two key equations are the Burgers formula~\citep{Burgers} for the displacement, 
and the Peach-Koehler formula~\citep{PK} for the stress. These equations are very 
important for the interaction between complex arrays of dislocations.
The classical 
description of the elastic fields produced by dislocations is based on
the theory of linear elasticity.
There is the problem
of mathematical singularities at the dislocation core, and 
an arbitrary core-cutoff radius which must be introduced to avoid divergence. 
In the classical continuum theory 
of dislocations~\citep{Kroener58,deWit60,Nabarro,Lardner,Mura,Teodosiu,Li} 
the concept of Volterra dislocations is used, and 
the dislocation core is described by a Dirac delta function.
This unsatisfactory situation can only be remedied, when 
the fact that physical dislocations have a finite core region and no
singularities exist are taken into account. 

As already pointed out by~\citet{Kroener58} and 
\citet{Lothe} the divergence can be avoided when 
dislocation distributions other than the delta function are used.
\citet{Lothe} considered a standard core model with constant density of dislocations in a planar strip with width $d$. However, the value of $d$ remains 
undetermined and the expressions for the elastic fields are more complicated
than their singular counterparts and difficult to use for generally
curved dislocations.
Moreover, for non-planar configurations the theory becomes much more complex.

In order to remove the singularities of dislocations and to model
the dislocation core more realistically, continuum theories of 
generalized elasticity may be used. 
A very promising candidate of such a theory is 
the so-called gradient elasticity theory.
The theory of gradient elasticity was originally proposed
by~\citet{Mindlin64,Mindlin65}, 
and \citet{Mindlin68} (see also~\citet{Eshel}). 
The correspondence between the strain gradient theory and the atomic structure
of materials with the nearest and next nearest interatomic interactions was
exhibited by~\citet{Toupin}.
The original Mindlin theory possesses too many new material parameters.
For isotropic materials,
Mindlin's theory of first strain gradient elasticity~\citep{Mindlin64,Mindlin68}
involves two characteristic lengths, and Mindlin's 
theory of second strain gradient elasticity~\citep{Mindlin65}
possesses four characteristic lengths.
The discrete nature of materials is inherently incorporated in the formulations 
through the characteristic lengths.
The capability of strain gradient theories in capturing size effects
is a direct manifestation of the involvement of characteristic lengths.
Simplified versions, which are  particular cases of 
Mindlin's theories, were proposed and used for dislocation modelling.
Such simplified gradient elasticity theories are known 
as gradient elasticity of Helmholtz type~\citep{LM05}, with only one material 
length scale parameter and 
gradient elasticity of bi-Helmholtz type~\citep{LM06,Lazaretal2006}
which involves two material length scale parameters as new
material coefficients.
Gradient elasticity is a continuum model of dislocations with 
core spreading.
Non-singular fields of straight dislocations 
were obtained in the framework of gradient elasticity of Helmholtz type
by~\citet{GA99,LM05,LM06,LMA05} and \citet{Gutkin00,Gutkin06} 
(see also,~\citet{Gutkin04}).
Surprisingly enough up until now, not a single work has been done in the direction of 
non-singular dislocation loops using strain gradient elasticity theory.
The reason may be in the expected mathematical complexity of the problem.
Such non-singular solutions of arbitrary dislocation
loops could be very useful for the so-called 
discrete dislocation dynamics (e.g.,~\citet{Li,Sun}).

The aim of this paper is to present non-singular 
solutions of arbitrary dislocation loops, 
by using simplified gradient elasticity theories. 
We present the key-formulae of dislocations loops valid in the
framework of gradient elasticity,
and also reemphasize straight dislocations in gradient elasticity.
The technique of Green functions for the key-formulae is used,
and analytical closed-form solutions for the dislocation fields
are derived.

The paper is organized as follows. 
In Section~2, the fundamentals of gradient elasticity of 
Helmholtz type are given.  
Dislocation loops and straight dislocations are investigated.
In Section~3, the theory of 
gradient elasticity of bi-Helmholtz type is considered. 
Dislocation loops and straight dislocations will be examined in this
framework. 
In Section~4, the conclusions are given. All the mathematical and technical 
details are given in the Appendices.

\section{Gradient elasticity of Helmholtz type}

A straightforward framework to obtain non-singular fields of dislocations
is the so-called theory of gradient elasticity. 
A simplified theory of strain gradient elasticity 
is called gradient elasticity of Helmholtz type~\citep{LM05,LM06}.
This gradient elasticity of Helmholtz type is a 
particular gradient elasticity theory 
evolving from Mindlin's general gradient elasticity 
theory~\citep{Mindlin64,Mindlin68}. 
This theory is also known as dipolar gradient elasticity theory~\citep{G03}, 
simplified strain gradient elasticity theory~\citep{GM10a,GM10b}
and special gradient elasticity theory~\citep{AA97}.
The theory of gradient elasticity of Helmholtz type is the gradient version 
of Eringen's
theory of nonlocal elasticity of Helmholtz type~\citep{Eringen83,Eringen02}
which is well-established.

The strain energy density of such a simplified gradient elasticity theory  
for an isotropic, linearly elastic material has the form~\citep{LM05,GM10a}
\begin{align}
\label{W}
W=\frac{1}{2}\, C_{ijkl}\beta_{ij}\beta_{kl}
+\frac{1}{2}\, \ell^2 C_{ijkl}\pd_m \beta_{ij} \pd_m \beta_{kl}\, ,
\end{align}
where $C_{ijkl}$ is the tensor of elastic moduli with the symmetry properties
\begin{align}
C_{ijkl}=C_{klij}=C_{jikl}=C_{ijlk}\, 
\end{align}
and it reads for an isotropic material
\begin{align}
\label{C}
C_{ijkl}=\mu\big(
\delta_{ik}\delta_{jl}+\delta_{il}\delta_{jk}\big)
+\lambda\, \delta_{ij}\delta_{kl}\,,
\end{align}
where $\mu$ and $\lambda$ are the Lam\'e moduli.
$\beta_{ij}$ denotes the elastic distortion tensor.
If the elastic distortion tensor is incompatible, it can be decomposed
as follows
\begin{align}
\label{B}
\beta_{ij}=\pd_j u_i -\beta^\TP_{ij}\,,
\end{align}
where $u_i$ and $\beta^\TP_{ij}$ denote the displacement vector and the plastic
distortion tensor, respectively. In addition, $\ell$ is the material length scale 
parameter of gradient elasticity of Helmholtz type. 
For dislocations, $\ell$ is related to
the dislocation core radius and is proportional to a lattice parameter.
Due to the symmetry of $C_{ijkl}$, Eq.~(\ref{W}) is equivalent to
\begin{align}
\label{W2}
W=\frac{1}{2}\, C_{ijkl}e_{ij}e_{kl}
+\frac{1}{2}\, \ell^2 C_{ijkl}\pd_m e_{ij} \pd_m e_{kl}\, ,
\end{align}
where $e_{ij}=1/2(\beta_{ij}+\beta_{ji})$ is the elastic strain tensor.
The condition for non-negative strain energy density, $W\ge 0$, gives 
\begin{align}
\label{cond-l}
(2\mu+3\lambda)\ge 0\,, \qquad\mu\ge 0\,,\qquad\ell^2\ge 0\,.
\end{align}

The reason that the elastic and plastic distortion tensors are incompatible can be 
the presence of dislocations.  
Dislocations cause self-stresses that means stresses caused without 
the presence of body forces.
The dislocation density tensor is defined in terms of the elastic and plastic 
distortion tensors as follows (e.g.,~\citet{Kroener58})
\begin{align}
\label{DD-el}
\alpha_{ij}&=\epsilon_{jkl}\pd_k \beta_{il}\,\\
\label{DD-pl}
\alpha_{ij}&=-\epsilon_{jkl}\pd_k \beta^\TP_{il}\,
\end{align}
and it fulfills the Bianchi identity of dislocations
\begin{align}
\label{BI}
\pd_j \alpha_{ij}=0\,,
\end{align}
which means that dislocations do not end inside the body. 
Eq.~(\ref{BI}) is a `conservation' law and shows that dislocations are 
source-free fields.

From Eq.~(\ref{W}) it follows that the constitutive equations are
\begin{align}
\label{CR1}
\sigma_{ij}&=\frac{\pd W}{\pd \beta_{ij}}
=\frac{\pd W}{\pd e_{ij}}
=C_{ijkl}\beta_{kl}=C_{ijkl}e_{kl}\,,\\
\label{CR2}
\tau_{ijk}&=\frac{\pd W}{\pd \pd_k\beta_{ij}}
=\frac{\pd W}{\pd \pd_k e_{ij}}
=\ell^2\,C_{ijmn}\pd_k \beta_{mn}=\ell^2\pd_k \sigma_{ij}\,,
\end{align}
were $\sigma_{ij}$ are the components of the Cauchy stress tensor, 
$\tau_{ijk}$ are the components of the so-called double stress tensor. 
It can be seen that $\ell$ is the characteristic length scale for 
double stresses.
Using Eqs.~(\ref{CR1}) and (\ref{CR2}), Eq.~(\ref{W2}) 
can also be written as~\citep{LM05} 
\begin{align}
\label{W3}
W=\frac{1}{2}\, \sigma_{ij}e_{ij}
+\frac{1}{2}\, \ell^2 \pd_k \sigma_{ij} \pd_k e_{ij}\, .
\end{align}
The strain energy density~(\ref{W3}) exhibits the symmetry both in
$\sigma_{ij}$ and $e_{ij}$ and in $\pd_k \sigma_{ij}$ and $\pd_k e_{ij}$.

The total stress tensor is given as a combination of the Cauchy stress tensor
and the divergence of the double stress tensor 
\begin{align}
\label{T-stress}
\sigma^0_{ij}=\sigma_{ij}-\pd_k \tau_{ijk}
=(1-\ell^2\Delta)\sigma_{ij}\,
\end{align}
and it fulfills the equilibrium condition
for vanishing body forces
\begin{align}
\pd_j\sigma^0_{ij}=\pd_j (\sigma_{ij}-\pd_k \tau_{ijk})=0\, .
\end{align}
The stress tensor $\sigma^0_{ij}$ is called in the notation
of~\citet{Jaunzemis} 
the polarization of the Cauchy stress $\sigma_{ij}$.
Due to gradient elasticity of Helmholtz type, Eq.~(\ref{T-stress}) reduces to 
an inhomogeneous Helmholtz equation where the total stress tensor is
the inhomogeneous piece. As pointed out by~\citet{LM05,LM06}, the total 
stress tensor may be identified with the singular classical stress tensor.
This identifies that the inhomogeneous Helmholtz equation~(\ref{T-stress}) is in full agreement with the equation for the stress proposed 
by~\citet{Eringen83,Eringen02} in his theory of nonlocal elasticity of Helmholtz 
type.

As shown by~\citet{LM05,LM06} the following governing equations 
for the displacement vector, the elastic distortion tensor,
the dislocation density tensor, and the plastic distortion tensor 
can be derived in the framework of gradient elasticity of Helmholtz type
\begin{align}
\label{u-H}
&L\, u_i =u_i^0\,,\\
\label{B-H}
&L\, \beta_{ij} =\beta_{ij}^0\,,\\
\label{A-H}
&L\, \alpha_{ij} =\alpha_{ij}^0\,,\\
\label{BP-H}
&L\, \beta^\TP_{ij} =\beta_{ij}^{\TP,0}\,,
\end{align}
where 
\begin{align}
\label{L}
L=1-\ell^2\Delta 
\end{align}
is the Helmholtz operator. 
The singular fields $u_i^0$, $\beta_{ij}^0$, 
$\alpha_{ij}^0$ and $\beta_{ij}^{\TP,0}$ are the sources
of the inhomogeneous Helmholtz equations (\ref{u-H})--(\ref{BP-H}).
The Helmholtz equations (\ref{u-H}) and (\ref{B-H}) can be further reduced to
Helmholtz-Navier equations
\begin{align}
\label{u-L}
&L\, L_{ik} u_k =C_{ijkl}\pd_j \beta^{\TP,0}_{kl}\,,\\
\label{B-L}
&L\, L_{ik}\beta_{km} =-C_{ijkl}\epsilon_{mlr}\pd_j \alpha_{kr}^0\,,
\end{align}
where $L_{ik}=C_{ijkl}\pd_j\pd_l$ is the differential operator of the Navier 
equation.  
For an isotropic material, it reads
\begin{align}
L_{ik}=\mu\, \delta_{ik}\Delta+(\mu+\lambda)\, \pd_i\pd_k\,.
\end{align}
In Eqs.~(\ref{u-L}) and (\ref{B-L}) the sources
are now the plastic distortion $\beta^{\TP,0}_{kl}$ and the 
dislocation density $\alpha_{kr}^0$ known from classical elasticity.

The corresponding three-dimensional Green tensor of the Helmholtz-Navier equation 
is defined by
\begin{align}
\label{pde-HN}
&L\, L_{ik}\, G_{kj} =-\delta_{ij}\, \delta(\Bx-\Bx')
\end{align}
and is calculated as (see Eq.~(\ref{GT-NH3}))
\begin{align}
\label{G}
G_{ij}(R)=\frac{1}{16\pi\mu(1-\nu)}\, \Big[2(1-\nu)\delta_{ij}\Delta-
\pd_i\pd_j\Big] A(R)\,,
\end{align}
with
\begin{align}
\label{A}
A(R)=R+\frac{2\ell^2}{R}\,
\Big(1-\e^{-R/\ell}\Big)\, 
\end{align} 
and $R = |\Bx-\Bx'|$.
In the limit $\ell\rightarrow 0$, the three-dimensional Green tensor
of classical elasticity~\citep{Mura,Li} 
is recovered from Eqs.~(\ref{G}) and (\ref{A}). 
It is important to note that $A(R)$ can be written as the convolution of
$R$ and $G(R)$:
\begin{align}
A(R)=R*G(R)\,,
\end{align}
where $*$ denotes the spatial convolution and
$G$ is the three-dimensional Green function of the Helmholtz equation
\begin{align}
\label{pde-H}
L\, G =\delta(\Bx-\Bx')\,. 
\end{align}
It reads~\citep{Wl}
\begin{align}
\label{G-H}
G(R)=\frac{1}{4\pi \ell^2 R}\, \e^{-R/\ell}\,.
\end{align}
In addition, it holds
\begin{align}
\Delta\Delta\, R=-8\pi\, \delta(\Bx-\Bx')\,.
\end{align}
The function~(\ref{A}) fulfills the relations
\begin{align}
\label{A-LDD}
L\,\Delta\Delta\, A(R)&=-8\pi\, \delta(\Bx-\Bx')\,,\\
\label{A-DD}
\Delta\Delta\, A(R)&=-8\pi\, G(R)\,,\\
\label{A-L}
L\, A(R)&=R\,.
\end{align}
Thus, $A(R)$ is the Green function of Eq.~(\ref{A-LDD})
which is a Helmholtz-bi-Laplace equation.

Using Eqs.~(\ref{Aij}) and (\ref{Aii}) for the differentiation of 
Eq.~(\ref{G}), the explicit form of the
three-dimensional Green tensor of the Helmholtz-Navier equation is obtained 
\begin{align}
\label{G-exp}
G_{ij}(R)&=\frac{1}{16\pi\mu(1-\nu)}\, \bigg[
\frac{\delta_{ij}}{R}\bigg((3-4\nu)\Big(1-\e^{-R/\ell}\Big)
+\frac{1}{R^2}\Big(2\ell^2-\big(R^2+2\ell R+2\ell^2\big)\e^{-R/\ell}\Big)\bigg)
\nonumber\\
&\hspace{3cm}
+\frac{R_i R_j}{R^3}
\bigg(1-\frac{6\ell^2}{R^2}+\bigg(2+\frac{6\ell}{R}+\frac{6\ell^2}{R^2}\bigg)
\e^{-R/\ell}\bigg)\bigg]\,,
\end{align}
which is non-singular.
It is worth noting as a check, that Eq.~(\ref{G-exp}) is in agreement with 
the corresponding expressions derived by~\citet{Polyzos} and \citet{GM09}
using slightly different approaches.

Note, that the Green tensor~(\ref{G-exp}) gives the non-singular 
displacement field, $u_i=G_{ij}f_j$ ($f_j$ is the constant
value of the magnitude of the point force 
acting at the arbitrary position $\Bx'$ in an infinite body), 
of the 
Kelvin point force problem~(e.g.,~\citet{Gurtin,Mura,HI})
in the framework of gradient elasticity of Helmholtz type.
The original solution of a concentrated force in an infinite body in the
context of the classical continuum theory of elasticity was given 
by~\citet{Kelvin}.

\subsection{Dislocation loops}
In this subsection, the
characteristic fields of dislocation loops in the framework of
gradient elasticity theory of Helmholtz type are calculated.

For a general (non-planar or planar) dislocation loop $L$, the 
classical dislocation density and the plastic distortion
tensors are~ (e.g.,~\citet{deWit2,Kossecka74})
\begin{align}
\label{A0}
\alpha^0_{ij}&=b_i\, \delta_j(L)=b_i \oint_L \delta(\Bx-\Bx')\, \d L'_j\,,\\
\label{B0}
\beta^{\TP, 0}_{ij}&=-b_i\, \delta_j(S)=-b_i \int_S \delta(\Bx-\Bx')\, \d S'_j\,,
\end{align}
where $b_i$ is the Burgers vector of the dislocation line element $\d L'_j$ at
$\Bx'$ and $\d S'_j$ is the dislocation loop area.
The surface $S$ is the dislocation surface, which is a cap of the 
dislocation line $L$.
$\delta_j(L)$ is the Dirac delta function for a closed curve $L$ and
$\delta_j(S)$ is the Dirac delta function for a surface $S$ 
with boundary $L$.

The solution of Eq.~(\ref{A-H}) can be written as the following convolution
integral
\begin{align}
\label{A-grad0}
\alpha_{ij}=G*\alpha_{ij}^0
=b_i \oint_L G(R)\, \d L'_j\,,
\end{align}
where $G(R)$ denotes
the three-dimensional Green function of the Helmholtz equation given 
by Eq.~(\ref{G-H}). 
The explicit solution of the dislocation density tensor for a dislocation loop
in gradient elasticity is calculated as
\begin{align}
\label{A-grad}
\alpha_{ij}(\Bx)=\frac{b_i}{4\pi \ell^2 }\,\oint_L \frac{\e^{-R/\ell}}{R}\, \d L'_j
\,,
\end{align}
describing a spreading dislocation core distribution.
The plastic distortion tensor of a dislocation loop, which is the solution
of Eq.~(\ref{BP-H}), is given by the convolution integral
\begin{align}
\label{BP-grad0}
\beta^\TP_{ij}=G*\beta_{ij}^{\TP,0}
=-b_i \int_S G(R)\, \d S'_j\, .
\end{align}
It reads as
\begin{align}
\label{BP-grad}
\beta^\TP_{ij}(\Bx)=-\frac{b_i}{4\pi \ell^2 }\,\int_S \frac{\e^{-R/\ell}}{R}\, \d S'_j\, .
\end{align}
Substituting Eq.~(\ref{BP-grad}) in Eq.~(\ref{DD-pl}) and using 
the Stokes theorem, we obtain formula~(\ref{A-grad}).

Using the Green tensor~(\ref{G}), and after a straightforward calculation
all the generalizations of the Mura, Peach-Koehler, and Burgers formulae
towards gradient elasticity can be obtained.
Starting with the elastic distortion tensor of a dislocation loop,
the solution of Eq.~(\ref{B-L}) gives the representation as 
the following convolution integral
\begin{align}
\label{B-Mura0}
\beta_{im}(\Bx)=\int_{-\infty}^\infty \epsilon_{mnr} C_{jkln} G_{ij,k}(R) 
\alpha^0_{lr}(\Bx')\, \d V'\,,
\end{align}
where $G_{ij,k}=\pd_k G_{ij}$.
Substituting the classical dislocation density tensor
of a dislocation loop (\ref{A0}) and carrying out the integration of
the delta function, we find the modified Mura formula
valid in gradient elasticity 
\begin{align}
\label{B-Mura}
\beta_{im}(\Bx)=\oint_L \epsilon_{mnr} b_l C_{jkln} G_{ij,k}(R)\, \d L'_r\,.
\end{align}
Substitute Eqs.~(\ref{C}) and (\ref{G}) into Eq.~(\ref{B-Mura}) 
and obtain after rearranging terms
\begin{align}
\label{B-grad0}
\beta_{ij}(\Bx)&=\frac{1}{8\pi}\oint_L
\epsilon_{jnr}\Big[\big(b_i\pd_n  -b_n\pd_i + b_l \delta_{in}\pd_l \big)\Delta
+\frac{1}{1-\nu} \,\big(b_n \pd_i \Delta-b_l \pd_i\pd_n\pd_l\big)\Big] A(R)\, 
\d L'_r\, .
\end{align}
Using the identity
\begin{align}
\epsilon_{rjn}\big(b_i\pd_n  -b_n\pd_i)
&=\epsilon_{rjn}\epsilon_{kin}\epsilon_{kst}b_s  \pd_t
=(\delta_{rk}\delta_{ji}-\delta_{ri}\delta_{jk})\epsilon_{kst}b_s  \pd_t
=(\epsilon_{rst}\delta_{ij}-\epsilon_{jst}\delta_{ir}) b_s \pd_t
\nonumber\\
&=(\epsilon_{rkl}\delta_{ij}-\epsilon_{jkl}\delta_{ir}) b_k \pd_l
\end{align}
and the relation
\begin{align}
&\oint_L \epsilon_{rjn} (b_n\pd_l-b_l\pd_n)\pd_l\pd_i A(R)\,  \d L'_r
=\oint_L b_s (\epsilon_{jst} \pd_r-\epsilon_{rst}\pd_j) \pd_t\pd_i A(R)\,  \d L'_r
\nonumber\\
&=\oint_L \d \big(b_s \epsilon_{jst} \pd_t\pd_i A(R)\big)
-\oint_L b_s \epsilon_{rst}\pd_t \pd_j\pd_i A(R)\,  \d L'_r
=-\oint_L b_s \epsilon_{rst}\pd_t \pd_j\pd_i A(R)\,  \d L'_r
\nonumber\\
&=-\oint_L b_k \epsilon_{rkl}\pd_l \pd_j\pd_i A(R)\,  \d L'_r\,,
\end{align}
the non-singular elastic distortion (\ref{B-grad0}) of a dislocation loop 
becomes
\begin{align}
\label{B-grad}
\beta_{ij}(\Bx)&=-\frac{b_k}{8\pi}\oint_L
\Big[\big(\epsilon_{jkl}\delta_{ir}-\epsilon_{rkl}\delta_{ij}
+\epsilon_{rij}\delta_{kl}\big)\pd_l \Delta
+\frac{1}{1-\nu} \,\epsilon_{rkl}\pd_l\pd_i\pd_j \Big] A(R)\, 
\d L'_r\, .
\end{align}
This is the `Mura formula' for a dislocation loop in gradient elasticity.
It is important to note that 
if Eq.~(\ref{B-grad}) is substituted into (\ref{DD-el}) and the 
relation~(\ref{A-DD}) is used, the dislocation density of a 
dislocation loop~(\ref{A-grad}) is recovered.

The symmetric part of the elastic distortion tensor (\ref{B-grad}) 
gives the elastic strain tensor of a dislocation loop
\begin{align}
\label{E-grad}
e_{ij}(\Bx)&=-\frac{b_k}{8\pi}\oint_L
\Big[\Big(
\frac{1}{2}\, \epsilon_{jkl}\delta_{ir}
+\frac{1}{2}\, \epsilon_{ikl}\delta_{jr}
-\epsilon_{rkl}\delta_{ij}\Big)\pd_l \Delta
+\frac{1}{1-\nu} \,\epsilon_{rkl}\pd_l\pd_i\pd_j \Big] A(R)\, 
\d L'_r\, .
\end{align}
The elastic dilatation of a dislocation loop is nothing but the trace of
the elastic strain~(\ref{E-grad})
\begin{align}
\label{Dil-grad}
e_{ii}(\Bx)&=\frac{(1-2\nu)b_k}{8\pi(1-\nu)}\oint_L
\epsilon_{rkl}\,\pd_l \Delta A(R)\, 
\d L'_r\, .
\end{align}
The elastic rotation vector is defined as the skewsymmetric part of the 
elastic distortion tensor $\omega_l=\frac{1}{2}\epsilon_{ijl}\beta_{ij}$
and reads
\begin{align}
\label{omega-grad}
\omega_{l}(\Bx)&=-\frac{b_k}{8\pi}\oint_L
\Big(\delta_{lr}\pd_k-\frac{1}{2}\, \delta_{kr}\pd_l\Big)\Delta  A(R)\, 
\d L'_r\, .
\end{align}
Using the constitutive relation~(\ref{CR1}) with Eq.~(\ref{C}), 
the non-singular stress field produced by a dislocation loop is found 
\begin{align}
\label{T-grad}
\sigma_{ij}(\Bx)&=-\frac{\mu b_k}{8\pi}\oint_L
\Big[\big(\epsilon_{jkl}\delta_{ir}
+\epsilon_{ikl}\delta_{jr}\big)\pd_l \Delta
+\frac{2}{1-\nu}\, \epsilon_{rkl}\big(\pd_i\pd_j-\delta_{ij}\Delta\big)\pd_l
\Big] A(R)\, \d L'_r\, ,
\end{align}
which can be interpreted as the Peach-Koehler formula within the framework of 
gradient elasticity.
One may verify that the stress is divergence-less, $\pd_j\sigma_{ij}=0$.
The double stress tensor of a dislocation loop is easily obtained 
if Eq.~(\ref{T-grad}) is substituted into Eq.~(\ref{CR2}).

The solution of Eq.~(\ref{u-L}) is the following convolution integral
\begin{align}
\label{u-Mura0}
u_i(\Bx)=-\int_{-\infty}^\infty C_{jkln} G_{ij,k}(R) \beta_{ln}^{\TP,0}(\xx')\,\d V'\, .
\end{align}
Substituting the classical plastic distortion 
of a dislocation loop (\ref{B0}) into Eq.~(\ref{u-Mura0})
gives the modified Volterra formula
valid in gradient elasticity 
\begin{align}
\label{u-Mura}
u_i(\Bx)=\int_S b_l C_{jkln} G_{ij,k}(R)\, \d S'_n\, .
\end{align}
Substituting Eqs.~(\ref{C}) and (\ref{G}) into Eq.~(\ref{u-Mura}) and rearranging
terms yield
\begin{align}
\label{u-Burgers-0}
u_i(\Bx)=\frac{b_l}{8\pi}\int_S 
\Big[\big(\delta_{il}\pd_n +\delta_{in}\pd_l -\delta_{ln}\pd_i\big)\Delta
+\frac{1}{1-\nu}\big(\delta_{ln}\Delta - \pd_l\pd_n\big)\pd_i \Big] 
A(R)\, \d S'_n\,.
\end{align}
Except the first term of Eq.~(\ref{u-Burgers-0}), we apply the Stokes theorem
in order to obtain line integrals
with 
\begin{align}
\int_S (\delta_{in}\pd_l-\delta_{ln}\pd_i)\Delta A(R)\, \d S'_n
=-\oint_L \epsilon_{ril}\Delta A(R)\, \d L'_r
\end{align}
and
\begin{align}
\int_S (\delta_{nl}\Delta-\pd_{l}\pd_n)\pd_i  A(R)\, \d S'_n
=-\oint_L \epsilon_{rlj}\pd_j\pd_i  A(R)\, \d L'_r\, .
\end{align}
In this way, the key-formula for the non-singular displacement vector 
in gradient elasticity is found 
\begin{align}
\label{u-Burger-grad}
u_i(\Bx) = \frac{b_i}{8\pi}\, \int_S \Delta\pd_j A(R)\, \d S'_j
+\frac{b_l\epsilon_{rlj}}{8\pi}\, \oint_L
\bigg\{\delta_{ij} \Delta -\frac{1}{1-\nu}\, \pd_i \pd_j   
\bigg\}\, A(R)\,  \d L'_r\, ,
\end{align}                  
which is the Burgers formula in the framework of gradient elasticity of
Helmholtz type.                         
Eq.~(\ref{u-Burger-grad}) determines the displacement field of a single 
dislocation loop.
The Eqs.~(\ref{B-grad})--(\ref{u-Burger-grad}) are straightforward, simple, and
closely resemble the singular solutions of classical elasticity theory.
In the limit $\ell\rightarrow 0$, the classical expressions are recovered
in Eqs.~(\ref{B-grad})--(\ref{u-Burger-grad}).
The expressions~(\ref{B-grad}), (\ref{T-grad}), and (\ref{u-Burger-grad}) 
retain most of the
analytic structure of the classical Mura, Peach-Koehler, and Burgers formulae.
The expressions~(\ref{B-grad})--(\ref{u-Burger-grad}) are
given in terms of the elementary function $A(R)$ given in Eq.~(\ref{A}), 
instead of the classical expression $R$.
The explicit expressions can be obtained by simple substitution of the
formulae for the derivatives of $A$ given in Eqs.~(\ref{Ai})--(\ref{Aiik}).
It is important to note that Eqs.~(\ref{B-grad})--(\ref{u-Burger-grad})
are non-singular due to the regularization of the classical singular
expressions (see Appendix A).
As an example, we substitute Eqs.~(\ref{Aijk}) and (\ref{Aiik}) into
Eq.~(\ref{T-grad}) and obtain the explicit expression for
the stress tensor
\begin{align}
\label{T-grad2}
\sigma_{ij}(\Bx)&=-\frac{\mu b_l}{8\pi}\oint_L
\bigg[\Big(\epsilon_{jkl}\delta_{ir}
+\epsilon_{ikl}\delta_{jr}
-\frac{2}{1-\nu}\,\epsilon_{rkl}\delta_{ij}\Big)
\frac{2R_k}{R^3}\Big[1-\Big(1+\frac{R}{\ell}\Big)\e^{-R/\ell}\Big]
\nonumber\\
&\qquad
+\frac{2}{1-\nu}\, \epsilon_{rkl}\bigg(
\frac{\delta_{ij}\,R_k+\delta_{ik}\, R_j+\delta_{jk}\, R_i}{R^3}\,
\Big[1-\frac{6\ell^2}{R^2}\,
\Big(1-\e^{-R/\ell}\Big)
+\Big(2+\frac{6\ell}{R}\Big)\,\e^{-R/\ell}\Big]\nonumber\\
&\qquad\quad
-\frac{3 R_iR_j R_k}{R^5}\,
\Big[1-\frac{10\ell^2}{R^2}\,
\Big(1-\e^{-R/\ell}\Big)
+\Big(4+\frac{10\ell}{R}+\frac{2R}{3\ell}\Big)\,\e^{-R/\ell}\Big]\bigg)
\bigg] \d L'_r\, .
\end{align}

To give the expression~(\ref{u-Burger-grad}) more
explicitly.
Using Eq.~(\ref{Aiik})
we introduce a generalized solid angle valid in gradient elasticity
of Helmholtz type
\begin{align}
\label{Omega}
\Omega(\Bx,\ell)=-\frac{1}{2}\, \int_S \Delta\pd_j A(R)\, \d S'_j
=\int_S\frac{R_j}{R^3}\Big(1-\Big(1+\frac{R}{\ell}\Big)\e^{-R/\ell}\Big)\, \d S'_j
\,.
\end{align}
Eq.~(\ref{Omega}) is non-singular and depends on the 
length scale $\ell$.
In the limit $\ell\rightarrow 0$, the usual solid angle (e.g.,~\citet{Li})
is recovered.
Thus, using Eq.~(\ref{Omega}) and carrying out some differentiations
with Eqs.~(\ref{Ai}) and (\ref{Aii}), we obtain from Eq.~(\ref{u-Burger-grad})
the explicit gradient elasticity version of the Burgers formula
\begin{align}
\label{u-Burger-grad-2}
u_i(\Bx) = &-\frac{b_i}{4\pi}\, \Omega(\Bx,\ell)
-\frac{b_l}{4\pi} \,\oint_L \epsilon_{ilr}\,
\frac{1}{R}\,\Big(1-\e^{-R/\ell}\Big)\d L'_r
\nonumber\\
&
-\frac{b_l}{8\pi (1-\nu)}
\oint_L \epsilon_{ljr}\pd_i\, 
\frac{R_j}{R}\Big(1-\frac{2\ell^2}{R^2}\,
\Big(1-\e^{-R/\ell}\Big)+\frac{2\ell}{R}\,\e^{-R/\ell}\Big)
\,  \d L'_r\, .
\end{align}    
The simplicity of our results is based on the use of gradient elasticity
theory of Helmholtz type.
Our results can be used in computer simulations of dislocation cores
at nano-scale
and in numerics as fast numerical sums of the relevant elastic fields
as it is used for the classical equations (e.g.,~\citet{Sun}).

\subsection{Straight dislocations}
In this subsection, using the modified Mura equation of gradient elasticity
of Helmholtz type~(\ref{B-Mura0}), the non-singular elastic
distortion fields of straight dislocations as a check of our general approach
are calculated.

\subsubsection{Screw dislocation}
A screw dislocation corresponds to the anti-plane strain problem. 
The Green function of the anti-plane strain problem in gradient elasticity
of Helmholtz type is nothing but the Green function of the
two-dimensional Helmholtz-Laplace equation and it reads (see Eq.~(\ref{GF-HL2}))
\begin{align}
\label{Gzz}
G_{zz}(R)=-\frac{1}{2\pi\mu}\Big\{\gamma_E+\ln R+K_0\big(R/\ell\big)\Big\}\,,
\end{align}
where $R=\sqrt{(x-x')^2+(y-y')^2}$,
$\gamma_E$ is the Euler constant and $K_n$ is the modified Bessel function of order $n$.
The Green function~(\ref{Gzz}) is non-singular.
The gradient of the Green function~(\ref{Gzz}) is obtained as
\begin{align}
\label{Gzzk}
G_{zz,k}(R)=-\frac{1}{2\pi\mu}\, \frac{R_k}{R^2}
\Big\{1-\frac{R}{\ell}\,K_1\big(R/\ell\big)\Big\}\,. 
\end{align}
Next, substituting Eq.~(\ref{Gzzk}) and the 
dislocation density of a screw dislocation 
$\alpha^0_{zz}=b_z\, \delta(x)\delta(y)$ into Eq.~(\ref{B-Mura0}), the 
elastic distortion produced by a screw dislocation is obtained.
For an infinite screw dislocation along the $z$-axis with Burgers vector 
$b_z$, the  
non-singular components for the elastic distortion are calculated as
\begin{align}
\label{dist-zx}
\beta_{zx}&=-\frac{b_z}{2\pi}\,\frac{y}{r^2}\Big\{1-\frac{r}{\ell}\, K_1(r/\ell )\Big\}\, ,\\
\label{dist-zy}
\beta_{zy}&=\frac{b_z}{2\pi}\,\frac{x}{r^2}\Big\{1-\frac{r}{\ell}\, K_1(r/\ell )\Big\}\, ,
\end{align}
where $r=\sqrt{x^2+y^2}$.
The expressions obtained earlier by~\citet{Lazar03} and
\citet{LM06} are recovered.
In the limit $\ell\rightarrow 0$, the classical expressions given by~\citet{deWit3}
are recovered in Eqs.~(\ref{dist-zx}) and (\ref{dist-zy}).

The Green function~(\ref{Gzz}) gives the non-singular 
displacement field 
$u_z=-G_{zz} f_z$ of a line force with the magnitude $f_z$ calculated 
by~\citet{LM06b} in the framework of gradient elasticity.

\subsubsection{Edge dislocation}
Now the plane strain problem of an edge dislocation is investigated.
The Green tensor of the plane strain problem in gradient elasticity
of Helmholtz type is derived as (see Eq.~(\ref{GT-NH2}))
\begin{align}
\label{G2d}
G_{ij}(R)=-&\frac{1}{2\pi\mu}\, \delta_{ij}
\Big\{\gamma_E+\ln R+K_0\big(R/\ell\big)\Big\}\nonumber\\
+&\frac{1}{16\pi\mu(1-\nu)}\, \pd_{i}\pd_j
\Big\{R^2\big(\gamma_E+\ln R\big)
+4\ell^2\big(\gamma_E +\ln R+K_0\big(R/\ell\big)\big)\Big\}\,. 
\end{align}
It is obvious that the terms proportional to the Euler constant do
not contribute to the elastic distortion fields.
The two-dimensional Green tensor~(\ref{G2d}) is non-singular.
In the limit $\ell\rightarrow 0$, the two-dimensional Green tensor
of classical elasticity~\citep{Mura,Li} 
is recovered in Eq.~(\ref{G2d}). 
The gradient of the Green tensor~(\ref{G2d}) is given by
\begin{align}
\label{GTk}
G_{ij,k}(R)&=
-\frac{1}{8\pi\mu(1-\nu)}\bigg[(3-4\nu)\, \delta_{ij}\, \frac{R_k}{R^2}
-\delta_{ik}\,\frac{R_j}{R^2}-\delta_{jk}\,\frac{R_i}{R^2}
+2\,\frac{R_i R_j R_k}{R^4}
\nonumber\\
&\qquad\qquad
+\frac{2}{R^2}\Big(\delta_{ij} R_k
+\delta_{ik}R_j+\delta_{jk}R_i
-4\,\frac{R_i R_j R_k}{R^2}\Big)
\Big(\frac{2\ell^2}{R^2}-K_2\big(R/\ell\big)\Big)
\nonumber\\
&\qquad\qquad
-\Big(
4(1-\nu)\, \delta_{ij}\, \frac{R_k}{\ell\, R}
-2\,\frac{R_i R_j R_k}{\ell\, R^3}\Big)K_1\big(R/\ell\big)\bigg]\, .
\end{align}
Substituting Eq.~(\ref{GTk}) and the 
dislocation density of an edge dislocation along $z$ axis with
Burgers vector $b_x$,
$\alpha^0_{xz}=b_x\, \delta(x)\delta(y)$,  into Eq.~(\ref{B-Mura0}),
the elastic  distortion of an edge dislocation is obtained.
Eventually, the non-vanishing components of the elastic distortion 
of an edge dislocation are calculated  as
\begin{align}
\label{dist-xx2}
\beta_{xx}&=-\frac{b_x}{4\pi(1-\nu)}\, 
\frac{y}{r^2}\,
\Big\{(1-2\nu)+\frac{2x^2}{r^2}+\frac{4\ell^2}{r^4}\,(y^2-3x^2)
-\frac{2(y^2-3x^2)}{r^2}\,K_2(r/\ell)\nonumber\\
&\hspace{6.5cm}
-\frac{2(y^2-\nu r^2)}{\ell  r }\, K_1(r/\ell)
\Big\} ,\\
\label{dist_xy2}
\beta_{xy}&=\frac{b_x}{4\pi(1-\nu)}\, 
\frac{x}{r^2}\,
\Big\{(3-2\nu)-\frac{2y^2}{r^2}
-\frac{4\ell^2}{r^4}\,(x^2-3 y^2)
+\frac{2(x^2-3 y^2)}{r^2}\,K_2(r/\ell)\nonumber\\
&\hspace{6.0cm}
-\frac{2\big(y^2+(1-\nu) r^2\big)}{\ell r}\,K_1(r/\ell)
\Big\},\\
\label{dist-yx2}
\beta_{yx}&=-\frac{b_x}{4\pi(1-\nu)}\, 
\frac{x}{r^2}\,
\Big\{(1-2\nu)+\frac{2y^2}{r^2}
+\frac{4\ell^2}{r^4}\,(x^2-3 y^2)
-\frac{2(x^2-3 y^2)}{r^2}\,K_2(r/\ell)\nonumber\\
&\hspace{6.0cm}
+\frac{2\big(y^2-(1-\nu) r^2\big)}{\ell r}\,K_1(r/\ell)
\Big\},\\
\label{dist-yy2}
\beta_{yy}&=-\frac{b_x}{4\pi(1-\nu)}\, 
\frac{y}{r^2}\,
\Big\{(1-2\nu)-\frac{2x^2}{r^2}-\frac{4\ell^2}{r^4}\,(y^2-3x^2)
+\frac{2(y^2-3x^2)}{r^2}\,K_2(r/\ell)\nonumber\\
&\hspace{6.5cm} 
-\frac{2(x^2-\nu r^2)}{\ell r}\,K_1(r/\ell)\Big\} ,
\end{align}
which are non-singular and agree 
with the formulae given by~\citet{Lazar03} and \citet{LM06}.
In the limit $\ell\rightarrow 0$, we obtain in 
Eqs.~(\ref{dist-xx2})--(\ref{dist-yy2}) 
the classical expressions given by~\citet{deWit3}.
As discussed by~\citet{Lazaretal2006}, the dislocation core radius can be defined 
straightforwardly in the framework of gradient elasticity as $R_c\simeq 6\, \ell$.
If $\ell\simeq 0.4\, a$, where $a$ denotes the lattice parameter,
is adopted as proposed by~\citet{Eringen83},
the dislocation core radius is $R_c\simeq 2.5\, a$. 
Using $\ell\simeq 0.4\, a$, the internal length reduces to $\ell\simeq 1.97$
\AA\ for lead (Pb) with $a=4.95$ \AA.

Note that the two-dimensional Green function~(\ref{G2d}) gives the non-singular 
displacement field, 
$u_i=-G_{ij} f_j$, of a line force with magnitude $f_j$ calculated 
by~\citet{LM06b} in the framework of gradient elasticity.

\section{Gradient elasticity of bi-Helmholtz type}
In this section, gradient elasticity theory of higher order is considered.
Gradient elasticity theory of higher order was originally introduced
by~\citet{Mindlin65,Mindlin72} (see also,~\citet{Jaunzemis,Wu,AL09}). 
Mindlin's theory of second strain gradient elasticity 
involves for isotropic materials, 
in addition to the two Lam\'e constants, sixteen additional material constants.
These constants produce four characteristic length scales.

A simple and robust gradient elasticity of higher order which
is called gradient elasticity theory of bi-Helmholtz type
was introduced by~\citet{Lazaretal2006} and \citet{LM06} 
and successfully applied to the problems of 
straight dislocations~\citep{Lazaretal2006,LM06},
straight disclinations~\citep{Deng2007} and point defects~\citep{Zhang}.
\citet{Lazaretal2006} and \citet{LM06} have shown that all state quantities
are non-singular.
By means of this second order gradient theory
it is possible to eliminate not only
the singularities of the strain and stress tensors, 
but also the singularities of the double and triple stress tensors and 
of the dislocation density tensors of straight dislocations 
at the dislocation line. 
In general, all fields
calculated in the theory of gradient elasticity of bi-Helmholtz type
are smoother than those calculated by gradient elasticity theory of Helmholtz type.
In general, there a two main motivations for the use of gradient elasticity 
of bi-Helmholtz type: 
a consistent regularization of all state quantities, 
and a more realistic modelling of dispersion relations.
A simple higher-order gradient theory 
in order to investigate dislocation loops should be used.
The theory of gradient elasticity of bi-Helmholtz type is the gradient
version of nonlocal elasticity of bi-Helmholtz type~\citep{LMA06}.

The strain energy density of gradient elasticity theory of bi-Helmholtz type
for an isotropic, linearly elastic material 
has the form~\citep{Lazaretal2006}
\begin{align}
\label{W-BH}
W=\frac{1}{2}\, C_{ijkl}\beta_{ij}\beta_{kl}
+\frac{1}{2}\, \ell_1^2 C_{ijkl}\pd_m \beta_{ij} \pd_m \beta_{kl}
+\frac{1}{2}\, \ell_2^4 C_{ijkl}\pd_n \pd_m \beta_{ij} \pd_n \pd_m \beta_{kl}
\,,
\end{align}
where $\ell_1=\ell$, $\ell_2$ is another characteristic length scale
and $C_{ijkl}$ is given in~(\ref{C}).
Due to the symmetry of $C_{ijkl}$, Eq.~(\ref{W-BH}) is equivalent to
\begin{align}
\label{W-BH2}
W=\frac{1}{2}\, C_{ijkl}e_{ij}e_{kl}
+\frac{1}{2}\, \ell_1^2 C_{ijkl}\pd_m e_{ij} \pd_m e_{kl}
+\frac{1}{2}\, \ell_2^4 C_{ijkl}\pd_n \pd_m e_{ij} \pd_n \pd_m e_{kl}
\, .
\end{align}
In addition to the constitutive equations~(\ref{CR1}) and (\ref{CR2})
another one is present in such a higher-order gradient theory,
\begin{align}
\label{CR3}
\tau_{ijkl}&=\frac{\pd W}{\pd \pd_l \pd_k\beta_{ij}}
=\frac{\pd W}{\pd \pd_l \pd_k e_{ij}}
=\ell_2^4\,C_{ijmn}\pd_l\pd_k \beta_{mn}=\ell_2^4\pd_l \pd_k \sigma_{ij}\,,
\end{align}
where $\tau_{ijkl}$ is called the triple stress tensor.
It can be seen that $\ell_2$ is the characteristic length scale for 
triple stresses. 
On the other hand, $\ell_1$ is the characteristic length scale for 
double stresses. 
Using Eqs.~(\ref{CR1}), (\ref{CR2}), and (\ref{CR3}), Eq.~(\ref{W-BH2}) 
can also be written as~\citep{Lazaretal2006}
\begin{align}
\label{W-BH3}
W=\frac{1}{2}\, \sigma_{ij}e_{ij}
+\frac{1}{2}\, \ell_1^2 \pd_k \sigma_{ij} \pd_k e_{ij}
+\frac{1}{2}\, \ell_2^4 \pd_l \pd_k \sigma_{ij} \pd_l \pd_k e_{ij}
\, .
\end{align}
The strain energy density~(\ref{W-BH3}) exhibits the symmetry in
$\sigma_{ij}$ and $e_{ij}$, in $\pd_k \sigma_{ij}$ and $\pd_k e_{ij}$, 
and in $\pd_l \pd_k \sigma_{ij}$ and $\pd_l \pd_k e_{ij}$.
The condition for non-negative strain energy density, $W\ge 0$, gives
\begin{align}
\label{vobd-l2}
\ell_1^2\ge 0\,,\qquad\ell_2^4\ge 0\,,
\end{align}
in addition to $(3\mu+2\lambda)\ge 0$ and $\mu\ge0$.

The total stress tensor reads now
\begin{align}
\label{T-stress-BH}
\sigma^0_{ij}=\sigma_{ij}-\pd_k \tau_{ijk}+\pd_l\pd_k \tau_{ijkl}\,.
\end{align}
In absence of body forces, the equation of equilibrium has the following
form
\begin{align}
\label{Eq-BH}
\pd_j\sigma^0_{ij}=
\pd_j(\sigma_{ij}-\pd_k \tau_{ijk}+\pd_l\pd_k \tau_{ijkl})=0\,.
\end{align}
Using Eqs.~(\ref{CR2}) and (\ref{CR3}), the total stress
tensor~(\ref{T-stress-BH}) can be written
\begin{align}
\label{T-stress-BH2}
\sigma^0_{ij}
=L \,\sigma_{ij}\,,
\end{align}
where the differential operator $L$ is given 
by
\begin{align}
\label{L-BH}
L=\big(1-\ell_1^2\Delta+\ell_2^4\Delta\Delta\big)
=\big(1-c_1^2\Delta\big)\big(1-c_2^2\Delta\big)
\end{align}
with
\begin{align}
\label{c1}
c^{2}_1&=\frac{\ell_1^{2}}{2}\bigg(1+\sqrt{1-4\,\frac{\ell_2^{4}}{\ell_1^{4}}}\bigg)\,,\\
\label{c2}
c^{2}_2&=\frac{\ell_1^{2}}{2}\bigg(1-\sqrt{1-4\,\frac{\ell_2^{4}}{\ell_1^{4}}}\bigg)
\end{align}
and 
\begin{align}
\ell_1^{2}&=c_1^{2}+c_2^{2}\, ,\\
\label{c1c2}
\ell_2^{4}&=c_1^{2}\, c_2^{2}\,.
\end{align}
Due to its structure as a product of two Helmholtz operators,
the differential operator~(\ref{L-BH}) is called bi-Helmholtz operator.

An important point, is the question concerning 
the mathematical character of the 
length scales $c_1$ and $c_2$.
\citet{Mindlin65} (see also,~\citet{Mindlin72,Wu}) pointed out that the conditions for non-negative 
$W$ supply no indications of the character, real or complex, of the 
characteristic lengths. 
\citet{Mindlin65} and \citet{Wu} 
have treated the characteristic lengths as if they were real and positive.
They also pointed out that a complex character of the lengths 
is equally admissible. 
The character, real or complex, of the lengths dictates the behaviour 
of the field variables.
In the theory of gradient elasticity of bi-Helmholtz type 
the condition for the character, real or complex, of the length scales
$c_1$ and $c_2$ can be obtained from the condition 
if the argument of the 
square root in Eqs.~(\ref{c1}) and (\ref{c2}) is positive or negative.
Thus, $c_1$ and $c_2$ are real if
\begin{align}
\label{Cond-c}
\ell_1^4-4\ell_2^4\ge 0\,,
\end{align}
and 
$c_1$ and $c_2$ are complex if
\begin{align}
\label{Cond-c2}
\ell_1^4-4\ell_2^4< 0\,.
\end{align}
If the lengths $c_1$ and $c_2$ are complex, then the behaviour of 
the solutions of the field quantities would be oscillatory.
In this case, the far-field behaviour of the strain and stress fields of
dislocations would not agree with the classical behaviour. 
The limit from gradient elasticity of bi-Helmholtz type to 
gradient elasticity of Helmholtz type is:
$c_2\rightarrow 0$, $\ell_2\rightarrow 0$ and $c_1\rightarrow \ell_1$.
If $c_1$ is complex, then also $\ell_1$ becomes complex 
what would be rather strange. 
Thus, a real character of the length scales $c_1$ and $c_2$ seems 
to be more realistic and more physical. 
In addition,  
\citet{Zhang} determined, in an atomistic calculation, 
the length scales $c_1$ and $c_2$ as positive and real for graphene.
In what follows, the length scales $c_1$ and $c_2$ will be treated
as if they are real and positive.

The Green tensor of the bi-Helmholtz-Navier equation is calculated as
(see Eq.~(\ref{GT-BH3}))
\begin{align}
\label{GT-BH}
G_{ij}(R)=\frac{1}{16\pi\mu(1-\nu)}\, \Big[2(1-\nu)\delta_{ij}\Delta-
\pd_i\pd_j\Big] A(R)\,,
\end{align}
where the elementary function~(\ref{A}) is changed to 
\begin{align}
\label{A-BH}
A(R)=R+\frac{2(c_1^2+c_2^2)}{R}
-\frac{2}{c_1^2-c_2^2}\,\frac{1}{R}
\Big(c_1^4\,\e^{-R/c_1}-c_2^4\,\e^{-R/c_2}\Big)\,.
\end{align}
Eq.~(\ref{A-BH}) is the Green function of the three-dimensional
bi-Helmholtz-bi-Laplace equation.
It is worth noting that the Green tensor~(\ref{GT-BH}) with
(\ref{A-BH}) is in agreement with the corresponding 
expression derived by~\citet{Zhang}.
On the other hand, 
the Green function of the bi-Helmholtz equation is given by (e.g.,~\citet{LMA06})
\begin{align}
\label{G-BH}
G(R)=\frac{1}{4\pi (c_1^2-c_2^2) R}\, 
\Big(\e^{-R/c_1}-\e^{-R/c_2}\Big)\,.
\end{align}
In the framework of gradient elasticity of bi-Helmholtz, 
the differential 
operator of bi-Helmholtz type (\ref{L-BH}) appears
in Eqs.~(\ref{u-H})--(\ref{BP-H}), (\ref{pde-HN}), (\ref{pde-H}), (\ref{A-LDD}) and (\ref{A-L}).

If we use Eqs.~(\ref{Aij-BH}) and (\ref{Aii-BH}) for the differentiation of 
Eq.~(\ref{GT-BH}), we obtain the explicit form of the
three-dimensional Green tensor of the bi-Helmholtz-Navier equation
\begin{align}
\label{G-BH-exp}
G_{ij}(R)&=\frac{1}{16\pi\mu(1-\nu)}\, \bigg[
\frac{\delta_{ij}}{R}\bigg((3-4\nu)
\bigg(1
-\frac{1}{c_1^2-c_2^2}\,
\Big(c_1^2\,\e^{-R/c_1}-c_2^2\,\e^{-R/c_2}\Big)
\bigg)
\nonumber\\
&\qquad\qquad
+\frac{2(c_1^2+c_2^2)}{R^2}
-\frac{2}{c_1^2-c_2^2}\,\frac{1}{R^2}
\Big(c_1^4\,\e^{-R/c_1}-c_2^4\,\e^{-R/c_2}\Big)
\nonumber\\
&\qquad\qquad
-\frac{2}{c_1^2-c_2^2}\,\frac{1}{R}
\Big(c_1^3\,\e^{-R/c_1}-c_2^3\,\e^{-R/c_2}\Big)
-\frac{1}{c_1^2-c_2^2}\,
\Big(c_1^2\,\e^{-R/c_1}-c_2^2\,\e^{-R/c_2}\Big)
\bigg)
\nonumber\\
&\quad
+\frac{R_i R_j}{R^3}
\bigg(
1-\frac{6(c_1^2+c_2^2)}{R^2}
+\frac{6}{c_1^2-c_2^2}\,\frac{1}{R^2}
\Big(c_1^4\,\e^{-R/c_1}-c_2^4\,\e^{-R/c_2}\Big)
\nonumber\\
&\qquad\qquad
+\frac{6}{c_1^2-c_2^2}\,\frac{1}{R}
\Big(c_1^3\,\e^{-R/c_1}-c_2^3\,\e^{-R/c_2}\Big)
+\frac{2}{c_1^2-c_2^2}\,
\Big(c_1^2\,\e^{-R/c_1}-c_2^2\,\e^{-R/c_2}\Big)
\bigg)\bigg]\, .
\end{align}

The Green tensor~(\ref{G-BH-exp}) 
gives the non-singular 
displacement field $u_i=G_{ij}f_j$ of the Kelvin point force problem, 
in the framework of gradient elasticity of bi-Helmholtz type.

\subsection{Dislocation loops}
The calculation of the characteristic fields of a dislocation loop in
gradient elasticity of bi-Helmholtz type, 
is analogous to the technique used in gradient elasticity of Helmholtz type.
The only difference in the results is that now 
the Green function~(\ref{G-BH}) and the elementary function~(\ref{A-BH}) 
of bi-Helmholtz type
enter the characteristic fields of a dislocation loop.
In gradient elasticity of bi-Helmholtz type, the dislocation density 
tensor~(\ref{A-grad0}) and the plastic distortion tensor~(\ref{BP-grad0}) 
are given in terms of the Green function of bi-Helmholtz type~(\ref{G-BH}). 
Thus, they are calculated as
\begin{align}
\label{A-grad-BH}
\alpha_{ij}(\Bx)
&=\frac{b_i}{4\pi (c_1^2-c_2^2) }\,\oint_L 
\frac{\e^{-R/c_1}-\e^{-R/c_2}}{R}\, \d L'_j
\,,\\
\label{BP-grad-BH}
\beta^\TP_{ij}(\Bx)&=
-\frac{b_i}{4\pi (c_1^2-c_2^2) }\,\int_S \frac{\e^{-R/c_1}-\e^{-R/c_2}}{R}\, \d S'_j\, .
\end{align}
In the limit $R\rightarrow 0$, the integrands of Eqs.~(\ref{A-grad-BH}) and
(\ref{BP-grad-BH}) are non-singular at the dislocation line 
in contrast to the corresponding ones,
Eqs.~(\ref{A-grad}) and (\ref{BP-grad}), calculated in gradient elasticity of Helmholtz type.   
On the other hand,
the elastic distortion tensor~(\ref{B-grad}), 
the elastic strain tensor~(\ref{E-grad}), 
the elastic dilatation~(\ref{Dil-grad}),
the elastic rotation vector~(\ref{omega-grad}),
the stress tensor~(\ref{T-grad}), and the displacement vector~(\ref{u-Burger-grad}) are given in terms of the elementary 
function~(\ref{A-BH}) and only (\ref{A-BH}) has to be substituted in these 
formulae.
The explicit formulae are not reproduced.
The only difference between the fields of a dislocation loop in 
gradient elasticity of bi-Helmholtz type, and of Helmholtz type is that 
the Green function of bi-Helmholtz type~(\ref{G-BH}) 
and the elementary function~(\ref{A-BH}) 
have to be substituted instead of the Green function of Helmholtz type~(\ref{G-H}) 
and the elementary function~(\ref{A-H}). 
For the derivatives of the function~(\ref{A-BH}), 
Eqs.~(\ref{A-BH2})--(\ref{Aiik-BH}) can be substituted into the corresponding 
formulae.
The characteristic fields of a dislocation loop in gradient elasticity of
bi-Helmholtz type retain all the analytical tensor structure of the 
corresponding classical formulae.

The triple stress tensor of a dislocation loop is easily obtained 
if the stress tensor $\sigma_{ij}$ is substituted into Eq.~(\ref{CR3}).
In gradient elasticity of bi-Helmholtz type the fields produced
by a dislocation loop are smoother that those predicted by gradient
elasticity of Helmholtz type.

\subsection{Straight dislocations}
In this subsection, the modified Mura equation~(\ref{B-Mura0}) is used for
gradient elasticity of bi-Helmholtz type. 
The technique of Green functions is used in order to
determine the non-singular elastic distortion of straight dislocations.

\subsubsection{Screw dislocation}
The Green function of the anti-plane strain problem in gradient elasticity
of bi-Helmholtz type is the Green function of the
two-dimensional bi-Helmholtz-Laplace equation and is given by (see Eq.~(\ref{GF-BHL2}))
\begin{align}
\label{Gzz-BH}
G_{zz}(R)=-\frac{1}{2\pi\mu}\Big\{\gamma_E+\ln R+
\frac{1}{c_1^2-c_2^2}\big[c_1^2 K_0\big(R/c_1\big)-c_2^2 K_0\big(R/c_2\big)\big]
\Big\}\,,
\end{align}
where $R=\sqrt{(x-x')^2+(y-y')^2}$.
The gradient of the Green function~(\ref{Gzz-BH}) is calculated as
\begin{align}
\label{Gzzk-BH}
G_{zz,k}(R)=-\frac{1}{2\pi\mu}\, \frac{R_k}{R^2}
\Big\{1-\frac{1}{c_1^2-c_2^2}\big[
c_1 R\,K_1\big(R/c_1\big)-c_2 R\,K_1\big(R/c_2\big]\Big\}\,. 
\end{align}
If Eq.~(\ref{Gzzk-BH}) and $\alpha^0_{zz}=b_z\, \delta(x)\delta(y)$ 
are substituted into Eq.~(\ref{B-Mura0}), the 
elastic distortion produced by a screw dislocation with Burgers vector $b_z$
is obtained
\begin{align}
\label{B-zx-BH}
\beta_{zx}&=-\frac{b_z}{2\pi}\,\frac{y}{r^2}
\Big\{1-
\frac{1}{c_1^2-c_2^2}\big[c_1 r\, K_1(r/c_1 )-c_2 r\,  K_1(r/c_2)\big]\Big\}\,,\\
\label{B-zy-BH}
\beta_{zy}&=\frac{b_z}{2\pi}\,\frac{x}{r^2}
\Big\{1-
\frac{1}{c_1^2-c_2^2}\big[c_1 r\, K_1(r/c_1 )-c_2 r\,  K_1(r/c_2)\big]\Big\}\,,
\end{align}
where $r=\sqrt{x^2+y^2}$.
Eqs.~(\ref{B-zx-BH}) and (\ref{B-zy-BH}) are in agreement with the expressions 
obtained by~\citet{LM06}.

The Green function~(\ref{Gzz-BH}) gives the non-singular 
displacement field 
$u_z=-G_{zz} f_z$ of a line force with the magnitude $f_z$ 
in the framework of gradient elasticity of bi-Helmholtz type.

\subsubsection{Edge dislocation}
The plane strain problem of an edge dislocation is now investigated.
The Green tensor of the plane strain problem in gradient elasticity
of bi-Helmholtz type is found as (see Eq.~(\ref{GT-BH2}))
\begin{align}
\label{G2d-BH}
G_{ij}(R)=-&\frac{1}{2\pi\mu}\, \delta_{ij}
\Big\{\gamma_E+\ln R+
\frac{1}{c_1^2-c_2^2}\big[c_1^2 K_0\big(R/c_1\big)-c_2^2 K_0\big(R/c_2\big)\big]\Big\}\nonumber\\
+&\frac{1}{16\pi\mu(1-\nu)}\, \pd_{i}\pd_j
\Big\{R^2\big(\gamma_E+\ln R\big)
+4\big(c_1^2+c_2^2\big)\big(\gamma_E +\ln R\big)\nonumber\\
&\qquad
+\frac{4}{c_1^2-c_2^2}\Big[c_1^4 K_0\big(R/c_1\big)-c_2^4 K_0\big(R/c_2\big)\Big]
\Big\}\,. 
\end{align}
The gradient of the Green tensor Eq.~(\ref{G2d-BH}) is calculated as
\begin{align}
\label{GTk-BH}
&G_{ij,k}(R)=
-\frac{1}{8\pi\mu(1-\nu)}\bigg[(3-4\nu)\, \delta_{ij}\, \frac{R_k}{R^2}
-\delta_{ik}\,\frac{R_j}{R^2}-\delta_{jk}\,\frac{R_i}{R^2}
+2\,\frac{R_i R_j R_k}{R^4}
\nonumber\\
&\ 
+\frac{2}{R^2}\Big(\delta_{ij} R_k
+\delta_{ik}R_j+\delta_{jk}R_i
-4\,\frac{R_i R_j R_k}{R^2}\Big)
\Big(\frac{2(c_1^2+c_2^2)}{R^2}
-\frac{1}{c_1^2-c_2^2}\Big[c_1^2 K_2\big(R/c_1\big)-c_2^2 K_2\big(R/c_2\big)\Big]
\Big)
\nonumber\\
&\hspace{2cm}
-\Big(
4(1-\nu)\, \delta_{ij}\, \frac{R_k}{ R}
-2\,\frac{R_i R_j R_k}{R^3}\Big)
\frac{1}{c_1^2-c_2^2}\Big[c_1 K_1\big(R/c_1\big)-c_2 K_1\big(R/c_2\big)\Big]
\bigg]\, .
\end{align}
If Eq.~(\ref{GTk-BH}) and $\alpha^0_{xz}=b_x\,\delta(x)\delta(y)$ 
are substituted into Eq.~(\ref{B-Mura0}),
the non-vanishing components of the elastic distortion 
of an edge dislocation are found as
\begin{align}
\label{B-xx-BH}
\beta_{xx}&=-\frac{b_x}{4\pi(1-\nu)}\, 
\frac{y}{r^2}\,
\Big\{(1-2\nu)+\frac{2x^2}{r^2}+\frac{4(c_1^2+c_2^2)}{r^4}\,(y^2-3x^2)
\nonumber\\
&\hspace{4cm}
-\frac{2(y^2-\nu r^2)}{r^2(c_1^2-c_2^2)}\,
\big[c_1 r\,K_1(r/c_1)-c_2 r\, K_1(r/c_2)\big]\nonumber\\
&\hspace{4cm}
-\frac{2(y^2-3x^2)}{(c_1^2-c_2^2) r^2}\,
\big[c_1^2\,K_2(r/c_1)-c_2^2\, K_2(r/c_2)\big]\Big\}\, ,\\
\label{B-xy-BH}
\beta_{xy}&=\frac{b_x}{4\pi(1-\nu)}\, 
\frac{x}{r^2}\,
\Big\{(3-2\nu)-\frac{2y^2}{r^2}
-\frac{4(c_1^2+c_2^2)}{r^4}\,(x^2-3 y^2)\nonumber\\
&\hspace{4cm}
-\frac{2\big(y^2+(1-\nu) r^2\big)}{(c_1^2-c_2^2)\, r^2}\,
\big[c_1 r K_1(r/c_1)-c_2 r  K_1(r/c_2)\big]
\nonumber\\
&\hspace{4cm}
+\frac{2(x^2-3 y^2)}{(c_1^2-c_2^2)\,r^2}\,
\big[c_1^2 K_2(r/c_1)-c_2^2  K_2(r/c_2)\big]\Big\}\,,\\
\label{B-yx-BH}
\beta_{yx}&=-\frac{b_x}{4\pi(1-\nu)}\, 
\frac{x}{r^2}\,
\Big\{(1-2\nu)+\frac{2y^2}{r^2}
+\frac{4(c_1^2+c_2^2)}{r^4}\,(x^2-3 y^2)\nonumber\\
&\hspace{4cm}
+\frac{2\big(y^2-(1-\nu) r^2\big)}{(c_1^2-c_2^2)\, r^2}\,
\big[c_1 r K_1(r/c_1)-c_2 r K_1(r/c_2)\big]
\nonumber\\
&\hspace{4cm}
-\frac{2(x^2-3 y^2)}{(c_1^2-c_2^2)\,r^2}\,
\big[c_1^2 K_2(r/c_1)-c_2^2 K_2(r/c_2)\big]\Big\}\,,\\
\label{B-yy-BH}
\beta_{yy}&=-\frac{b_x}{4\pi(1-\nu)}\, 
\frac{y}{r^2}\,
\Big\{(1-2\nu)-\frac{2x^2}{r^2}-\frac{4(c_1^2+c_2^2)}{r^4}\,(y^2-3x^2)\nonumber\\
&\hspace{4cm}
-\frac{2(x^2-\nu r^2)}{r^2(c_1^2-c_2^2)}\,
\big[c_1 r\,K_1(r/c_1)-c_2 r\, K_1(r/c_2)\big]
\nonumber\\
&\hspace{4cm}
+\frac{2(y^2-3x^2)}{(c_1^2-c_2^2) r^2}\,
\big[c_1^2\,K_2(r/c_1)-c_2^2\, K_2(r/c_2)\big]\Big\}\, ,
\end{align}
which are in agreement
with the formulae given by~\citet{LM06}.

The two-dimensional Green function~(\ref{G2d-BH}) gives the non-singular 
displacement field, 
$u_i=-G_{ij} f_j$, of a line force with magnitude $f_j$ calculated 
in the framework of gradient elasticity of bi-Helmholtz type.

\section{Conclusions}
Non-singular dislocation fields are presented 
in the framework of gradient elasticity. 
The technique of Green functions is used. 
The Green tensors
of all relevant partial differential equations of generalized 
Navier type were calculated.
For the first time, the elastic distortion, plastic 
distortion, stress, displacement, and dislocation density 
of a closed dislocation loop, using the theories of 
gradient elasticity of Helmholtz type and 
of bi-Helmholtz type were calculated. 
Straight dislocations using Green tensors were revisited.
Such generalized continuum theories allow 
dislocation core spreading in a straightforward manner.
In classical dislocation theory the dislocation function is a Dirac delta function, 
$\delta(\xx)$, without core spreading.
In the non-singular approaches by~\citet{Cai06} and  Lazar, presented in the
present paper, the 
dislocation spreading functions are $w$ and $G$, respectively
(see Table~\ref{table}).
In the theory of gradient elasticity all formulae are closed
in contrast to the theory of~\citet{Cai06} where the spreading function 
$w$ is determined in a sophisticated way in order to obtain
$R_a=[R^2+a^2]^{1/2}$.
Due to the use of simplified theories of gradient elasticity, 
the dislocation fields retain most of the analytical structure of the classical
expressions for these quantities but remove the singularity 
at the dislocation core due to the mathematical regularization of the
classical singular expressions.
In gradient elasticity of Helmholtz type, 
the characteristic length $\ell$ takes into account the information from
atomistic calculations 
as discussed in this paper.
In a straightforward manner, the length $\ell$ 
determines the dislocation core radius. Therefore, in gradient elasticity
it is not necessary to introduce an artifical core-cutoff radius.
It should be mentioned that 
the characteristic lengths 
which arise in first strain gradient elasticity~(e.g.,~\citet{Sharma07,Shodja10})
and in second strain gradient elasticity~(e.g.,~\citet{Zhang,Shodja12})
have been recently computed using atomistic approaches.

\begin{table}[t]
\caption{Comparison of the basic quantities in different dislocation theories 
(classical dislocation theory,
theory of Cai {\it et al.} and gradient theory of Helmholtz type)}

\begin{center}
\leavevmode
\begin{tabular}{||c|c|c||}\hline
Classical theory~\citep{deWit60} & \citet{Cai06} & Lazar~[this paper]\\
\hline
$R$  & $R_a$ & $A(R)$\\
  $\Delta\Delta\, R=-8\pi \delta(\xx)$ & 
 $\Delta\Delta\, R_a =-8 \pi w$ &  
$\Delta\Delta\, A(R)=-8 \pi G$ \\
& $R_a=R*w$ & $A(R)=R*G$ \\
& $w$ chosen to obtain: 
& $L\, G=\delta(\xx)$\,,\quad $L=1-\ell^2\Delta$ \\
&$R_a=[R^2+a^2]^{1/2}$& 
$A(R)=R+\frac{2\ell^2}{R}\big(1-\e^{-R/\ell}\big)$\\
& $a$ -- arbitrary constant & $\ell$ -- characteristic length\\
& $w=15 a^4/[8\pi (r^2+a^2)^{7/2}]$ 
& $G=\e^{-r/\ell}/[4\pi \ell^2 r]$\\
\hline
\end{tabular}
\end{center}
\label{table}
\end{table}

The obtained results can be used in computer simulations and numerics of 
dislocation cores, discrete dislocation dynamics, 
and arbitrary 3D dislocation configurations.
The results can be implemented in dislocation dynamics codes (finite element
implementation, technique of fast numerical sums), 
and compared to atomistic models (e.g.,~\citet{Sun,Li}). 

\section*{Acknowledgements}
The author gratefully acknowledges the grants from the 
Deutsche Forschungsgemeinschaft (Grant Nos. La1974/2-1, La1974/3-1). 

\begin{appendix}
\section{Appendix: $A$ and its derivatives}
\label{appendixA}
\setcounter{equation}{0}
\renewcommand{\theequation}{\thesection.\arabic{equation}}

In gradient elasticity theory, the stress tensor, the elastic distortion tensor,
the elastic strain tensor, and the displacement vector of a dislocation loop 
are given in terms of derivatives of the elementary function $A$.

\subsection{Helmholtz type}
For gradient elasticity of Helmholtz type, the elementary function $A$ is given by
\begin{align}
\label{A2}
A=R+\frac{2\ell^2}{R}\,
\Big(1-\e^{-R/\ell}\Big)\, .
\end{align} 
Higher-order derivatives of $A$ are given by the following set of equations
\begin{align}
\label{Ai}
A_{,i}=\frac{R_i}{R}\Big[1-\frac{2\ell^2}{R^2}\,
\Big(1-\e^{-R/\ell}\Big)+\frac{2\ell}{R}\,\e^{-R/\ell}\Big]
\, ,
\end{align} 
where $R_i=x_i-x'_i$,
\begin{align}
\label{Aij}
A_{,ij}=\frac{\delta_{ij}}{R}\Big[1-\frac{2\ell^2}{R^2}\,
\Big(1-\e^{-R/\ell}\Big)+\frac{2\ell}{R}\,\e^{-R/\ell}\Big]
-\frac{R_{i}R_j}{R^3}\Big[1-\frac{6\ell^2}{R^2}\,
\Big(1-\e^{-R/\ell}\Big)
+\Big(2+\frac{6\ell}{R}\Big)\,\e^{-R/\ell}\Big]
\, ,
\end{align} 
\begin{align}
\label{Aii}
A_{,ii}=\frac{2}{R}\big(1-\e^{-R/\ell}\big)\,,
\end{align} 
\begin{align}
\label{Aijk}
A_{,ijk}=&-\frac{\delta_{ij}\,R_k+\delta_{ik}\, R_j+\delta_{jk}\, R_i}{R^3}\,
\Big[1-\frac{6\ell^2}{R^2}\,
\Big(1-\e^{-R/\ell}\Big)
+\Big(2+\frac{6\ell}{R}\Big)\,\e^{-R/\ell}\Big]\nonumber\\
&+\frac{3 R_iR_j R_k}{R^5}\,
\Big[1-\frac{10\ell^2}{R^2}\,
\Big(1-\e^{-R/\ell}\Big)
+\Big(4+\frac{10\ell}{R}+\frac{2R}{3\ell}\Big)\,\e^{-R/\ell}\Big]
\, 
\end{align} 
and
\begin{align}
\label{Aiik}
A_{,iik}&=-\frac{2R_k}{R^3}\Big(1-\Big(1+\frac{R}{\ell}\Big)\e^{-R/\ell}\Big)
\,.
\end{align}
The expressions~(\ref{A2})--(\ref{Aiik}) are non-singular.
For $R\rightarrow 0$, they are either zero or finite.

\subsection{Bi-Helmholtz type}
In gradient elasticity of bi-Helmholtz type, 
the elementary function $A$ reads
\begin{align}
\label{A-BH2}
A=R+\frac{2(c_1^2+c_2^2)}{R}
-\frac{2}{c_1^2-c_2^2}\,\frac{1}{R}
\Big(c_1^4\,\e^{-R/c_1}-c_2^4\,\e^{-R/c_2}\Big)\,.
\end{align} 
The higher-order derivatives of $A$ are given by
\begin{align}
\label{Ai-BH}
A_{,i}=\frac{R_i}{R}\bigg[1-\frac{2(c_1^2+c_2^2)}{R^2}
+\frac{2}{c_1^2-c_2^2}\,\frac{1}{R^2}
\Big(c_1^4\,\e^{-R/c_1}-c_2^4\,\e^{-R/c_2}\Big)
+\frac{2}{c_1^2-c_2^2}\,\frac{1}{R}
\Big(c_1^3\,\e^{-R/c_1}-c_2^3\,\e^{-R/c_2}\Big)
\bigg]
\, ,
\end{align} 
\begin{align}
\label{Aij-BH}
A_{,ij}&=\frac{\delta_{ij}}{R}\bigg[1-\frac{2(c_1^2+c_2^2)}{R^2}
+\frac{2}{c_1^2-c_2^2}\,\frac{1}{R^2}
\Big(c_1^4\,\e^{-R/c_1}-c_2^4\,\e^{-R/c_2}\Big)
+\frac{2}{c_1^2-c_2^2}\,\frac{1}{R}
\Big(c_1^3\,\e^{-R/c_1}-c_2^3\,\e^{-R/c_2}\Big)
\bigg]\nonumber\\
&\ 
-\frac{R_{i}R_j}{R^3}\bigg[1-\frac{6(c_1^2+c_2^2)}{R^2}
+\frac{6}{c_1^2-c_2^2}\,\frac{1}{R^2}
\Big(c_1^4\,\e^{-R/c_1}-c_2^4\,\e^{-R/c_2}\Big)
+\frac{6}{c_1^2-c_2^2}\,\frac{1}{R}
\Big(c_1^3\,\e^{-R/c_1}-c_2^3\,\e^{-R/c_2}\Big)\nonumber\\
&\qquad\qquad
+\frac{2}{c_1^2-c_2^2}\,
\Big(c_1^2\,\e^{-R/c_1}-c_2^2\,\e^{-R/c_2}\Big)
\bigg]\, ,
\end{align} 
\begin{align}
\label{Aii-BH}
A_{,ii}=\frac{2}{R}\bigg[1
-\frac{1}{c_1^2-c_2^2}\,
\Big(c_1^2\,\e^{-R/c_1}-c_2^2\,\e^{-R/c_2}\Big)
\bigg]\,,
\end{align} 
\begin{align}
\label{Aijk-BH}
A_{,ijk}=&-\frac{\delta_{ij}\,R_k+\delta_{ik}\, R_j+\delta_{jk}\, R_i}{R^3}\,
\bigg[1-\frac{6(c_1^2+c_2^2)}{R^2}
+\frac{6}{c_1^2-c_2^2}\,\frac{1}{R^2}
\Big(c_1^4\,\e^{-R/c_1}-c_2^4\,\e^{-R/c_2}\Big)
\nonumber\\
&\qquad\qquad
+\frac{6}{c_1^2-c_2^2}\,\frac{1}{R}
\Big(c_1^3\,\e^{-R/c_1}-c_2^3\,\e^{-R/c_2}\Big)
+\frac{2}{c_1^2-c_2^2}
\Big(c_1^2\,\e^{-R/c_1}-c_2^2\,\e^{-R/c_2}\Big)
\bigg]\nonumber\\
&+\frac{3 R_iR_j R_k}{R^5}\,
\bigg[1-\frac{10(c_1^2+c_2^2)}{R^2}
+\frac{10}{c_1^2-c_2^2}\,\frac{1}{R^2}
\Big(c_1^4\,\e^{-R/c_1}-c_2^4\,\e^{-R/c_2}\Big)
\nonumber\\
&\qquad\qquad
+\frac{10}{c_1^2-c_2^2}\,\frac{1}{R}
\Big(c_1^3\,\e^{-R/c_1}-c_2^3\,\e^{-R/c_2}\Big)
+\frac{4}{c_1^2-c_2^2}
\Big(c_1^2\,\e^{-R/c_1}-c_2^2\,\e^{-R/c_2}\Big)
\nonumber\\
&\qquad\qquad\qquad
+\frac{2 R}{3(c_1^2-c_2^2)}
\Big(c_1\,\e^{-R/c_1}-c_2\,\e^{-R/c_2}\Big)
\bigg]
\, 
\end{align} 
and
\begin{align}
\label{Aiik-BH}
A_{,iik}&=-\frac{2R_k}{R^3}
\bigg[1-\frac{1}{c_1^2-c_2^2}\,
\Big(c_1^2\,\e^{-R/c_1}-c_2^2\,\e^{-R/c_2}\Big)
-\frac{R}{c_1^2-c_2^2}\,
\Big(c_1\,\e^{-R/c_1}-c_2\,\e^{-R/c_2}\Big)
\bigg]\,.
\end{align}
The expressions~(\ref{A-BH2})--(\ref{Aiik-BH}) are non-singular.
In the limit $c_2\rightarrow 0$ and $c_1=\ell$, 
Eqs.~(\ref{A-BH2})--(\ref{Aiik-BH}) reduce to
Eqs.~(\ref{A2})--(\ref{Aiik}).

\section{Appendix: Green tensors of generalized Navier equations}

\label{appendixB}
\setcounter{equation}{0}
\renewcommand{\theequation}{\thesection.\arabic{equation}}

The following notation is used for the $n$-dimensional Fourier transform \citep{GC} 
\begin{align}
\widetilde f(\kk)&\equiv \FF_{(n)}\big[f(\rr)\big]=
\int_{-\infty}^{+\infty} f(\rr)\, \e^{+\ii \kk\cdot\rr} \d\rr\, ,\\
f(\rr)&\equiv \FF^{-1}_{(n)}\big[\widetilde f(\kk)\big]=
\frac{1}{(2\pi)^n}\int_{-\infty}^{+\infty} \widetilde f(\kk)\, \e^{-\ii \kk\cdot\rr} \d\kk\, .
\end{align}
We have~\citep{Wl,Nowacki}
\begin{align}
\label{FT2-k2}
\FF^{-1}_{(2)}\left[\frac{1}{k^2}\right]&=
-\frac{1}{2\pi}\,\big( \gamma_E+\ln \sqrt{x^2+y^2}\big)\,, \\
\label{FT3-k2}
\FF^{-1}_{(3)}\left[\frac{1}{k^2}\right]&=
\frac{1}{4\pi}\, \frac{1}{\sqrt{x^2+y^2+z^2}}\,, \\
\label{FT2-k4}
\FF^{-1}_{(2)}\left[\frac{1}{k^4}\right]&=
\frac{1}{8\pi}\, (x^2+y^2)\big(\gamma_E+\ln \sqrt{x^2+y^2}\big)\,, \\
\label{FT3-k4}
\FF^{-1}_{(3)}\left[\frac{1}{k^4}\right]&=
-\frac{1}{8\pi}\, \sqrt{x^2+y^2+z^2}\,, \\
\label{FT-K2}
\FF^{-1}_{(2)}\left[\frac{1}{k^2+\frac{1}{c^2}}\right]&=
\frac{1}{2\pi}\, K_0\big( \sqrt{x^2+y^2}/c\big)\,, \\
\label{FT-K3}
\FF^{-1}_{(3)}\left[\frac{1}{k^2+\frac{1}{c^2}}\right]&=
\frac{1}{4\pi \sqrt{x^2+y^2+z^2}}\, \exp\big(-\sqrt{x^2+y^2+z^2}/c\big)\,.
\end{align}

\subsection{Green tensor of the Helmholtz-Navier 
equation}

The Green tensor of the Helmholtz-Navier equation is defined by
\begin{align}
\label{GT-x}
(1-\ell^2\Delta)\big(\mu\, \delta_{il}\Delta+(\lambda+\mu) \pd_i \pd_l\big){G}_{lj}(r)=-\delta_{ij}\delta(\Bx)\,.
\end{align}
The Fourier transform of Eq.~(\ref{GT-x}) reads
\begin{align}
\label{GT-k}
(1+\ell^2k^2)\big(\mu\, \delta_{il}k^2+(\lambda+\mu) k_i k_l\big)\widetilde{G}_{lj}(k)=\delta_{ij}\,,
\end{align}
where $\lambda=2\mu\nu/(1-2\nu)$ and $\nu$ is Poisson's ratio.
The Fourier transformed Green tensor  is found as
\begin{align}
\widetilde{G}_{ij}(k)=\frac{1}{\mu}\bigg[\frac{\delta_{ij}}{k^2}
-\frac{1}{2(1-\nu)}\,\frac{k_ik_j}{k^4}\bigg]\frac{1}{1+\ell^2k^2}\,.
\end{align}

Using partial fractions and the inverse Fourier transform, we find
\begin{align}
\FF^{-1}_{(3)}\Big(\frac{k_ik_j}{k^4(1+\ell^2k^2)}\Big)&=
-\pd_i\pd_j\, \FF^{-1}_{(3)}\Big(\frac{1}{k^4(1+\ell^2k^2)}\Big)\nonumber\\
&=-\pd_i\pd_j\, \FF^{-1}_{(3)}
\Big(\frac{1}{k^4}-\frac{\ell^2}{k^2}+\frac{\ell^2}{k^2+\frac{1}{\ell^2}}\Big)
\nonumber\\
&=\frac{\pd_i\pd_j}{8\pi}
\Big(r+\frac{2\ell^2}{r}-\frac{2\ell^2}{r}\, \e^{-r/\ell}\Big)
\end{align}
and
\begin{align}
\FF^{-1}_{(3)}\Big(\frac{1}{k^2(1+\ell^2k^2)}\Big)&=
-\Delta\, \FF^{-1}_{(3)}\Big(\frac{1}{k^4(1+\ell^2k^2)}\Big)\nonumber\\
&=\frac{\Delta}{8\pi}
\Big(r+\frac{2\ell^2}{r}-\frac{2\ell^2}{r}\, \e^{-r/\ell}\Big)\,,
\end{align}
the three-dimensional Green tensor of the Helmholtz-Navier equation 
is calculated as
\begin{align}
\label{GT-NH3}
G_{ij}(r)=\frac{1}{16\pi\mu(1-\nu)}\, \Big[2(1-\nu)\delta_{ij}\Delta-
\pd_i\pd_j\Big] \Big[r+\frac{2\ell^2}{r}\,
\Big(1-\e^{-r/\ell}\Big)\Big]\, ,
\end{align} 
where $r = \sqrt{x^2+y^2+z^2}$.

On the other hand, using
\begin{align}
\label{F1}
\FF^{-1}_{(2)}\Big(\frac{k_ik_j}{k^4(1+\ell^2k^2)}\Big)&=
-\pd_i\pd_j\, \FF^{-1}_{(2)}\Big(\frac{1}{k^4(1+\ell^2k^2)}\Big)\nonumber\\
&=-\pd_i\pd_j\, \FF^{-1}_{(2)}
\Big(\frac{1}{k^4}-\frac{\ell^2}{k^2}+\frac{\ell^2}{k^2+\frac{1}{\ell^2}}\Big)
\nonumber\\
&=-\frac{\pd_i\pd_j}{8\pi}
\Big(r^2(\gamma_E+\ln r) +4\ell^2\big(\gamma_E+\ln r)+K_0(r/\ell)\big)\Big)
\end{align}
and
\begin{align}
\label{F2}
\FF^{-1}_{(2)}\Big(\frac{1}{k^2(1+\ell^2k^2)}\Big)&=
\FF^{-1}_{(2)}\Big(\frac{1}{k^2}-\frac{1}{k^2+\frac{1}{\ell^2}}
\Big)\nonumber\\
&=-\frac{1}{2\pi}
\Big(\gamma_E+\ln r+K_0(r/\ell)\Big)\,,
\end{align}
the two-dimensional Green tensor of the Helmholtz-Navier equation 
is obtained as
\begin{align}
\label{GT-NH2}
G_{ij}(r)=-&\frac{1}{2\pi\mu}\, \delta_{ij}
\Big\{\gamma_E+\ln r+K_0\big(r/\ell\big)\Big\}\nonumber\\
+&\frac{1}{16\pi\mu(1-\nu)}\, \pd_{i}\pd_j
\Big\{r^2\big(\gamma_E+\ln r\big)
+4\ell^2\big(\gamma_E +\ln r+K_0\big(r/\ell\big)\big)\Big\}\,,
\end{align} 
where $r = \sqrt{x^2+y^2}$.

\subsection{Green function of the Helmholtz-Laplace
equation}

For the anti-plane strain problem,
the Green tensor of the Navier-Helmholtz equation 
reduces to the Green function of the two-dimensional Helmholtz-Laplace equation 
which is defined by
\begin{align}
\label{GF-HL}
(1-\ell^2\Delta)\Delta\, {G}_{zz}(r)=-\frac{1}{\mu}\,\delta(\Bx)\,.
\end{align}
The Fourier transform of Eq.~(\ref{GF-HL}) reads
\begin{align}
\label{GF-HL-k}
(1+\ell^2k^2) k^2 \widetilde{G}_{zz}(k)=\frac{1}{\mu}\, .
\end{align}
The Fourier transformed Green function is
\begin{align}
\widetilde{G}_{zz}(k)=\frac{1}{\mu}\,\frac{1}{k^2(1+\ell^2k^2)}\,.
\end{align}
Using Eq.~(\ref{F2}), the two-dimensional Green function 
is calculated as
\begin{align}
\label{GF-HL2}
G_{zz}(r)=-&\frac{1}{2\pi\mu}\, 
\Big\{\gamma_E+\ln r+K_0\big(r/\ell\big)\Big\}\,.
\end{align}

\subsection{Green tensor of the bi-Helmholtz-Navier 
equation}

The Green tensor of the bi-Helmholtz-Navier equation is defined by
\begin{align}
\label{GT-x-BH}
(1-c_1^2\Delta)(1-c_2^2\Delta)
\big(\mu\, \delta_{il}\Delta+(\lambda+\mu) \pd_i \pd_l\big){G}_{lj}(r)=-\delta_{ij}\delta(\Bx)\,.
\end{align}
The Fourier transform of Eq.~(\ref{GT-x-BH}) reads
\begin{align}
\label{GT-k-BH}
(1+c_1^2k^2)(1+c_2^2k^2)
\big(\mu\, \delta_{il}k^2+(\lambda+\mu) k_i k_l\big)\widetilde{G}_{lj}(k)=\delta_{ij}\,.
\end{align}
The Fourier space Green tensor  is 
\begin{align}
\widetilde{G}_{ij}(k)=\frac{1}{\mu}\bigg[\frac{\delta_{ij}}{k^2}
-\frac{1}{2(1-\nu)}\,\frac{k_ik_j}{k^4}\bigg]\frac{1}{(1+c_1^2k^2)
(1+c_2^2k^2)}\,.
\end{align}

Using
\begin{align}
\FF^{-1}_{(3)}\Big(\frac{k_ik_j}{k^4(1+c_1^2k^2)(1+c_2^2k^2)}\Big)&=
-\pd_i\pd_j\, \FF^{-1}_{(3)}\Big(\frac{1}{k^4(1+c_1^2k^2)(1+c_2^2k^2)
}\Big)\nonumber\\
&=-\pd_i\pd_j\, \FF^{-1}_{(3)}
\Big(\frac{1}{k^4}-\frac{c_1^2+c_2^2}{k^2}
+\frac{1}{c_1^2-c_2^2}\Big(
\frac{c_1^4}{k^2+\frac{1}{c_1^2}}
-\frac{c_2^4}{k^2+\frac{1}{c_2^2}}
\Big)\Big)
\nonumber\\
&=\frac{\pd_i\pd_j}{8\pi}
\Big(r+\frac{2(c_1^2+c_2^2)}{r}
-\frac{2}{c_1^2-c_2^2}\,\frac{1}{r}\Big(
c_1^4\, \e^{-r/c_1}-c_2^4\, \e^{-r/c_2}\Big)\Big)
\end{align}
and
\begin{align}
\FF^{-1}_{(3)}\Big(\frac{1}{k^2(1+c_1^2k^2)(1+c_2^2k^2)}\Big)&=
-\Delta\, \FF^{-1}_{(3)}\Big(\frac{1}{k^4(1+c_1^2k^2)(1+c_2^2k^2)}\Big)
\nonumber\\
&=\frac{\Delta}{8\pi}
\Big(r+\frac{2(c_1^2+c_2^2)}{r}
-\frac{2}{c_1^2-c_2^2}\,\frac{1}{r}\Big(
c_1^4\, \e^{-r/c_1}-c_2^4\, \e^{-r/c_2}\Big)\Big)
\,,
\end{align}
the three-dimensional Green tensor of the bi-Helmholtz-Navier equation 
is found as
\begin{align}
\label{GT-BH3}
G_{ij}(r)=\frac{1}{16\pi\mu(1-\nu)}\, \Big[2(1-\nu)\delta_{ij}\Delta-
\pd_i\pd_j\Big] \Big[r+\frac{2(c_1^2+c_2^2)}{r}
-\frac{2}{c_1^2-c_2^2}\,\frac{1}{r}\Big(
c_1^4\, \e^{-r/c_1}-c_2^4\, \e^{-r/c_2}\Big)\Big]\, ,
\end{align} 
where $r = \sqrt{x^2+y^2+z^2}$.

In two dimensions, we use the formulae
\begin{align}
\label{F3}
\FF^{-1}_{(2)}\Big(\frac{1}{k^2(1+c_1^2k^2)(1+c_2^2k^2)}\Big)
&=\FF^{-1}_{(2)}
\Big(\frac{1}{k^2}-\frac{1}{c_1^2-c_2^2}\Big(
\frac{c_1^2}{k^2+\frac{1}{c_1^2}}
-\frac{c_2^2}{k^2+\frac{1}{c_2^2}}
\Big)\Big)
\nonumber\\
&=-\frac{1}{4\pi}
\Big(\gamma_E+\ln r
+\frac{1}{c_1^2-c_2^2}\,\big[
c_1^2\, K_0(r/c_1)-c_2^2\, K_0(r/c_2)\big]\Big)
\end{align}
and
\begin{align}
\FF^{-1}_{(2)}\Big(\frac{k_ik_j}{k^4(1+c_1^2k^2)(1+c_2^2k^2)}\Big)&=
-\pd_i\pd_j\, \FF^{-1}_{(2)}\Big(\frac{1}{k^4(1+c_1^2k^2)(1+c_2^2k^2)
}\Big)\nonumber\\
&=-\pd_i\pd_j\, \FF^{-1}_{(2)}
\Big(\frac{1}{k^4}-\frac{c_1^2+c_2^2}{k^2}
+\frac{1}{c_1^2-c_2^2}\Big(
\frac{c_1^4}{k^2+\frac{1}{c_1^2}}
-\frac{c_2^4}{k^2+\frac{1}{c_2^2}}
\Big)\Big)
\nonumber\\
&=-\frac{\pd_i\pd_j}{8\pi}
\Big(r^2(\gamma_E+\ln r)+4(c_1^2+c_2^2)(\gamma_E+\ln r)\nonumber\\
&\qquad
+\frac{4}{c_1^2-c_2^2}\,\big[
c_1^4\, K_0(r/c_1)-c_2^4\, K_0(r/c_2)\big]\Big)\,.
\end{align}
Eventually, the two-dimensional Green tensor of the bi-Helmholtz-Navier equations 
is obtained as
\begin{align}
\label{GT-BH2}
G_{ij}(r)=-&\frac{1}{2\pi\mu}\, \delta_{ij}
\Big\{\gamma_E+\ln r+
\frac{1}{c_1^2-c_2^2}\big[c_1^2 K_0\big(r/c_1\big)-c_2^2 K_0\big(r/c_2\big)\big]\Big\}\nonumber\\
+&\frac{1}{16\pi\mu(1-\nu)}\, \pd_{i}\pd_j
\Big\{R^2\big(\gamma_E+\ln r\big)
+4\big(c_1^2+c_2^2\big)\big(\gamma_E +\ln r\big)\nonumber\\
&\qquad
+\frac{4}{c_1^2-c_2^2}\Big[c_1^4 K_0\big(r/c_1\big)-c_2^4 K_0\big(r/c_2\big)\Big]
\Big\}\,. 
\end{align}

\subsection{Green function of the bi-Helmholtz-Laplace
equation}
For the anti-plane strain problem,
the Green tensor of the Navier-Helmholtz equation 
reduces to the Green function of the two-dimensional bi-Helmholtz-Laplace equation 
which is defined by
\begin{align}
\label{GF-BHL}
(1-c_1^2\Delta)(1-c_2^2\Delta)
\Delta\, {G}_{zz}(r)=-\frac{1}{\mu}\,\delta(\Bx)\,.
\end{align}
The Fourier transform of Eq.~(\ref{GF-BHL}) reads
\begin{align}
\label{GF-BHL-k}
(1+c_1^2k^2)(1+c_2^2k^2) k^2 \widetilde{G}_{zz}(k)=\frac{1}{\mu}\, .
\end{align}
The Fourier transformed Green function is
\begin{align}
\widetilde{G}_{zz}(k)=
\frac{1}{\mu}\,\frac{1}{k^2(1+c_1^2k^2)(1+c_2^2k^2)}\,.
\end{align}
Using Eq.~(\ref{F3}), the two-dimensional Green function 
is obtained as
\begin{align}
\label{GF-BHL2}
G_{zz}(r)=-\frac{1}{2\pi\mu}\Big\{\gamma_E+\ln r+
\frac{1}{c_1^2-c_2^2}\big[c_1^2 K_0\big(r/c_1\big)-c_2^2 K_0\big(r/c_2\big)\big]
\Big\}\,.
\end{align}

\end{appendix}

\end{document}